\documentclass[aip,rsi,amsmath,amssymb,reprint]{revtex4-1}
\usepackage{graphicx}
\usepackage{bm}
\usepackage[font=footnotesize,centerlast]{caption}

\begin{document}

\title{Non-local Corrections to Collisional Transport in Magnetised Plasmas}

\author{H. C. Watkins}
\email{h.watkins@ucl.ac.uk}
\altaffiliation[Now at ]{University College London.}
\author{R. J. Kingham}
\affiliation{Blackett Laboratory, Imperial College London, London SW7 2AZ, United Kingdom
}

\date{\today}

\begin{abstract}
In modern inertial fusion experiments there is a complex interplay between non-locality and magnetisation that can greatly influence transport. In this work we use a matrix recursion method to include higher-order corrections beyond the diffusion approximation usually used for magnetised plasmas. Working in the linear regime, we show this can account for arbitrary orders of the distribution function expansion in Knudsen non-locality parameter $k\lambda_{ei}$. Transport coefficients, such as thermal conductivity, deviate from the magnetised diffusive approximation. In particular we show how higher orders of the expansion contribute to transport on a plasma perturbation asynchronously parallel, perpendicular and cross-perpendicular to a uniform magnetic field.
\end{abstract}

\maketitle

\section{Introduction}
\label{S:1}

ICF experiments consistently show a deviation in thermal transport from the local Spitzer-Harm approximation \cite{Farmer2018, Michel2015MeasurementsOMEGA, Gregori2004EffectPropagation, Albritton1983LaserDistributions,Albritton1986NonlocalDistributions}. The Nernst effect and the Righi-Leduc heat flow in magnetised simulations and experiments are similarly affected \cite{Joglekar2016KineticHohlraums, Kho1985NonlinearPlasmas,Kho1986NonlinearPlasmas}. As such there is a need for a model that can be used in codes that accurately reproduces the transport that is present in Vlasov-Fokker-Planck (VFP) simulations without the computational overhead of a kinetic code. 

Improvements in the accurate calculation and simulation of plasma physics phenomena is essential in laser-based fusion research. Reliable simulations are necessary to design experiments and to explain their results. Constructing an accurate non-local transport model relies on understanding how certain approximations contribute to the error. 

The conventional derivation of fluid theory relies on the truncation of the expansion of the electron distribution function in velocity space  \cite{Shkarofsky1966ThePlasmas}, $f_0+\textbf{f}_1\cdot\frac{\textbf{v}}{v}$. Here $f_0$  and $f_1$ are the zeroth-order and first order terms respectively in the Cartesian tensor  expansion of the distribution function. It is assumed there is only a small deviation of the distribution function away from a Maxwellian. This `diffusive approximation' relies on terms higher than $f_1$ being small enough to be ignored. A further approximation is made where it is assumed the isotropic part $f_0$ is itself Maxwellian.  

This local approximation holds in the regime $L_T>>\lambda_{ei}$, where the electron-ion collision mean-free-path $\lambda_{ei}$ is much smaller than the temperature scale length $L_T$. This condition however is frequently violated in high-intensity laser-plasma interactions \cite{Bell1981ElecronPlasmas,Bell1983ElectronPlasmas}. Simulations of plasmas (especially in hohlraum regimes in inertial confinement fusion (ICF) and scrape-off layers (SOL) in tokamaks) cannot accurately predict the heat flow due to this non-local nature.

Magnetic fields have been observed by proton radiography in hohlraum plasmas \cite{Manuel2012FirstPlasmas,Li2009ObservationsRadiography}. These self-induced fields appear as a result of phenomena such as the Biermann-battery effect. Magnetic fields can also be present in plasmas through external field coils. This latter method has been favoured in recent experiments \cite{Hohenberger2012, Montgomery2015UseLaser-coupling, Perkins2017TheFusion} to take advantage of the reduction in thermal transport under a magnetic field to improve the fusion yield. Simulations using Epplerain-Haines transport coefficients \cite{Epperlein1986PlasmaEquation} have also been performed that show the impact of magnetised transport on hohlraum conditions \cite{Farmer2017SimulationEnvironment}. The inclusion of a magnetic field acts as a localisation mechanism \cite{Froula2007LaserPlasmas}, and thus complicates the creation of an accurate model to simulate non-local transport by adding an anisotropic effective mean free path.

In this work we will be analysing the influence of the higher polynomial terms of the SH expansion on magnetised transport for a plasma in the Lorentz limit ($Z>>1$). These terms are ignored in local fluid models but also in conventional non-local transport models such as the SNB model \cite{Schurtz2000ACodes,Nicolai2006APlasmas}. We examine whether ignoring higher SH terms in the expansion accounts for the error in magnetised transport effects such as the Nernst effect and perpendicular thermal transport. To incorporate the higher order terms, our analysis will consider the linearised electron VFP model with a magnetic field. This will enable the use of a continued fraction method introduced by Epperlein \cite{Epperlein1992DampingCollisions}, used previously in local linear theories without a magnetic field \cite{Bychenkov1995NonlocalPlasma,Brantov2004KineticPlasmas}. We find a magnetised non-local correction to the $f_1$ equation that incorporates contributions from an arbitrary number of terms in the expansion of the electron distribution function. Thereby we extend the work of Brantov \cite{Brantov2003LinearPlasma} to go beyond the diffusive approximation.  This factor can be considered a modified collision frequency and can be used to calculate a non-local correction to the transport coefficients in the linear regime.

By investigating the system with a reduced model, we show how the truncation of the distribution function expansion leads to errors in the non-local, magnetised regime relative to the diffusive approximation. The corresponding transport coefficients show non-monotonic deviation of up to $50\%$ from the diffusive approximation at intermediate non-locality values ($k\lambda_{ei} \approx 10-100$). We also present the unexpected result of non-monotonic non-locality dependence of perpendicular transport coefficients, which has never before been reported. This leads to transport coefficients varying significantly from expected behaviour \cite{Brantov2003LinearPlasma}, and suggests a complex interplay between the localising effect of magnetic fields and the non-locality of non-Maxweillian distributions.

We begin by a brief discussion of non-local transport under magnetic fields in section \ref{S:2}. Section III will present the matrix recursion method and calculate the corrections needed to calculate the modified coefficients. In section \ref{S:4} will show how the corrected coefficients differ from the classical forms derived in the diffusive approximation. Finally, section \ref{S:5} will discuss how these results impact laser-plasma experiments and the parameter space where the corrections are important. The appendix \ref{S:Ap} presents the calculations of the transport coefficients in the classical diffusion approximation.  

\section{Non-local Magnetised Transport}
\label{S:2}

In the presence of sufficient Coulomb collisions, the fluid approximation assumes the distribution function of electrons in a plasma is approximately Maxwellian,
\begin{equation}
    f_m=n_e\left(\frac{1}{2\pi v_{th}^2}\right)^{3/2}\exp{\left(-\frac{v^2}{2 v_{th}^2}\right)}.
\end{equation}

However in the presence of steep temperature gradients, electrons with collisional mean-free paths greater than the scale length of the gradient (where $L_T<<\lambda_{ei}$), stream down the temperature gradient without thermalising to the local distribution. Equivalently, one can follow Epperlein \cite{Epperlein1994EffectFlow} and consider a wave mode of wavenumber $k$ and define a non-locality parameter, the Fourier-space Knudsen number, $\eta$ where 
\begin{equation}
    \eta=k\lambda_{ei}(v).
\end{equation}

Where one can define also a thermal value $\eta_{th} = k\lambda_{ei}(v_{th})$. Modes with high values of $\eta$ will be less collisional and the electrons unable to relax to be locally Maxwellian. The deviation from an isotropic Maxwellian motivates the use of a spherical harmonic expansion of the distribution function,
\begin{equation}
    f_e(\textbf{x},\textbf{v},t)=f_0+\textbf{f}_1\cdot\frac{\textbf{v}}{v}+\textbf{f}_2:\frac{\textbf{vv}}{v^2}+...
\end{equation}

The anisotropy in the distribution is contained in the terms $\textbf{f}_1$ and higher. The inherently anisotropic nature of magnetic fields means their influence is applied to these $f_n, n>1$ terms.

\subsection{Magnetised Form of the Electron VFP}
In the diffusion approximation the electron VFP equation becomes the two equations 
\begin{gather}
\frac{\partial f_0 }{\partial t}+\frac{v}{3}\nabla\cdot\textbf{f}_1+\frac{1}{3v^2}\frac{\partial}{\partial v}\left(v^2 \textbf{a}\cdot\textbf{f}_1 \right)=C_{e0}, \\
\frac{\partial \textbf{f}_1}{\partial t}+v\nabla f_0+\textbf{a}\frac{\partial f_0}{\partial v}+\bm{\omega}\times\textbf{f}_1 = -\nu_{ei}\textbf{f}_1 +\textbf{C}_{e1}.
\end{gather}
Electron-electron collisions are present through the terms $C_{e0}$ and $\textbf{C}_{e1}$ and where the electromagnetic fields are defined as
\begin{gather}
\textbf{a}=-\frac{e}{m_e}\textbf{E},\\
\bm{\omega}=-\frac{e}{m_e}\textbf{B}.
\end{gather}

Following Epperlein and Haines \cite{Epperlein1986PlasmaEquation},  time dependence is ignored in eq. 5 and so $f_1$ can be instantaneously calculated from $f_0$,
\begin{equation}
    \textbf{f}_1=-\frac{1}{\nu_{ei}(v)}\textbf{M}\left(v \nabla f_0+\textbf{a}\frac{\partial f_0}{\partial v}\right).
\end{equation}
This equation contains the matrix $\textbf{M}$ which is the inverse of $\textbf{I} + \frac{\bm{\omega}\times}{\nu_{ei}}$. The magnetic field has been incorporated into this `magnetisation matrix' defined as
\begin{equation}
M_{ij}=\frac{\chi_i \chi_j}{\chi^2}+\left(\delta_{ij}-\frac{\chi_i \chi_j}{\chi^2}\right)\frac{1}{1+\chi^2} -\frac{\epsilon_{ikj}\chi_k}{1+\chi^2},
\end{equation}
where $\epsilon_{ikj}$ is the Levi-Civita tensor. $\bm{\chi}$ is the Hall vector, itself defined in terms of the Larmor frequency $\bm{\omega}=-\frac{e B}{m_e}$ and the collision frequency $\nu_{ei}$ as
\begin{equation}
    \bm{\chi}=\frac{\bm{\omega}}{ \nu_{ei}(v)}.
\end{equation}

One can see the thermal value of this parameter is $\chi_{th}=\omega/\nu_{ei}(v_{th})$. The three terms of this matrix represent the fluxes parallel, perpendicular and cross-perpendicular to the magnetic field respectively. Given this definition of $\textbf{f}_1$, it is possible to calculate the fluxes of charge (current, $\textbf{j}$) and thermal energy (the intrinsic heat flow, $\textbf{q'}$), and in turn the transport coefficients using the moment expressions \cite{Swanson2008PlasmaTheory},
\begin{gather}
    \textbf{j}=-e\frac{4\pi}{3}\int^\infty_0 \textbf{f}_1 v^3 dv, \\
    \textbf{q}=\frac{4 \pi}{3} \frac{m_e}{2}\int^\infty_0 \textbf{f}_1 v^5 dv,\\
    \textbf{q'}=\textbf{q}+\frac{5 T_e}{4 e}\textbf{j}.
\end{gather}

This work will be considering the influence on linear perturbations to the electron fluid. We follow here the derivation used by Epperlein \cite{Epperlein1986PlasmaEquation, Epperlein1986ThePlasma}, albeit in a linearised model. A full calculation of the local transport coefficients in this manner can be found in Appendix \ref{S:Ap}. They will form the basis of the non-local corrected forms, as we shall take these local coefficients and modify them with correction factors that follow.   

By linearising the isotropic part $f_0$ about a Maxwellian distribution $f_0 = f_m + \delta f_0$, we can take the linear parts of eq. 8.  After performing a Fourier transform, the $\textbf{f}_1$ equation is
\begin{equation}
    \hat{\textbf{f}}_1=-\frac{1}{\nu_{ei}(v)} \textbf{M} \left[i \textbf{k}v \delta \hat{f}_0+\hat{\textbf{a}}\frac{\partial f_m}{\partial v}\right].
\end{equation}
Where the Maxwellian $f_m$ is defined in terms of the uniform background density and temperature $n_0, T_0$, with $v_{th}=\sqrt{T_0/m_e}$.

The transport coefficients are recovered by forming the Onsager transport relations from moments defined by eqs. 11-13. The Onsager form is then transformed into the classical transport coefficient forms. Full details of this procedure can be found in Appendix A.

\section{Calculating the High-Polynomial Corrections}
\label{S:3}

The methodology described above can be extended to incorporate the higher-order terms by finding a correction factor that sits in the matrix $\textbf{M}$, to find such a correction we look first at continued fractions. 

\subsection{Continued Fractions}

In the linear regime the Vlasov-Fokker-Planck equation can, using the SH expansion, be represented as an infinite set of coupled linear equations. The system can written in the form of a recurrence relation \cite{Epperlein1992DampingCollisions}, 
\begin{equation}
    f_{l+1}+b_l f_l + a_l f_{l-1}=0.    
\end{equation}

In the unmagnetised case, this recurrence relation has the corresponding continued fraction
\begin{equation}
    \frac{f_l}{f_{l-1}}=\cfrac{-a_l}{b_l+\cfrac{-a_{l+1}}{b_{l+1}+\cfrac{-a_{l+2}}{b_{l+2}+...}}}.
\end{equation}

Each successive equation in the hierarchy contributes to the fraction, which can be shown to converge at infinity. This form is used by Epperlein \cite{Epperlein1992DampingCollisions} to calculate a correction factor $H(\eta)$ to the collision frequency that encodes the dependence on non-locality ($\eta$). It is derived by `summing up' the contributions of the entire infinite hierarchy of linear equations, and he finds it to be approximately
\begin{equation}
    H(\eta)=\sqrt{1+\left(\frac{\pi \eta}{6}\right)^2}.
\end{equation}

By analogy with this unmagnetised result, the aim will be to find a modified magnetisation matrix $\textbf{G}$ that encapsulates the effect of the higher-order terms. In the limit of nonlocality going to zero, we expect to retrieve the unmodified magnetisation matrix.
\begin{equation}
\lim_{\eta\to 0} \textbf{G}(\eta) = \textbf{M}
\end{equation}

The matrix $\textbf{G}$ would then follow through to the calculation of the coefficients via the moment integrals in Appendix A. The question is how to calculate this $\textbf{G}$ matrix from the infinite hierarchy of linear equations that come out of the SH expansion of the magnetised VFP.

\subsection{Spherical Harmonics and the KALOS formalism}

To construct a recurrence relation that will apply to a magnetised plasma, we turn to the KALOS formalism. The KALOS formalism \cite{Bell2006} solves the electron Vlasov-Fokker-Planck equation by decomposing the distribution function into spherical harmonics, 
\begin{equation}
    f_e(\textbf{x},\textbf{v},t)=\sum^\infty_{l=0}\sum^l_{m=-l}f_l^m(\textbf{x},v,t)P^m_l(\cos \theta)e^{im\phi},
\end{equation}

and solving the resulting non-linear system of equations numerically to an arbitrary choice of order of harmonic. For our purposes we shall only need the specific form for a linear perturbation in one direction. After linearising, the KALOS system of equations become, for $l> 1$,
\begin{gather}
    \frac{\partial f_l^m}{\partial t}=A_l^m + B_l^m + C_l^m,\\
    C_l^m=-\frac{l(l+1)}{2}\nu_{ei} (v)f_l^m,\\
    A_l^m=-\left(\frac{l-m}{2l-1}\right)v\frac{\partial f_{l-1}^m}{\partial x}-\left(\frac{l+m+1}{2l+3}\right)v\frac{\partial f_{l+1}^m}{\partial x}.
\end{gather}

Following the analysis of Epperlein \cite{Epperlein1992DampingCollisions}, the electric field term $E_l^m$ will only appear in the $l=0$ equation. The collision term $C_l^m$ only contains contributions from e-i collisions, under the assumption the e-e collisions only contribute weakly to terms $l>0$ when in the regime of high ion charge. For the magnetic field term, there are two forms depending on the value of $m$, for $m=0$,
\begin{equation}
    \Re(B_l^0)=-\frac{e}{m_e}l(l+1)(B_z \Re(f_l^1)+B_y \Im(f_l^1)),
\end{equation}
and for $m>0$,

\begin{widetext}
\begin{equation}
    B_l^m=-i\frac{e B_x}{m_e}mf_l^m-\frac{e}{2 m_e}\left[(l-m)(l+m+1)(B_z-iB_y)f_l^{m+1}-(B_z+iB_y)f_l^{m-1}\right].
\end{equation}
\end{widetext}

Using a Fourier transform this set becomes an algebraic recurrence relation where each $f_l^m$ is coupled to $f_l^{m+1}$, $f_l^{m-1}$, $f_{l+1}^m$ and $f_{l-1}^m$.

With the x-axis set by the direction of the perturbation wavevector ($\hat{\textbf{x}}=\hat{\textbf{k}}$), we are free to define the z-direction as $\hat{\textbf{z}}=(\hat{\textbf{k}}\times \hat{\textbf{B}})\times\hat{\textbf{k}}$; in this case the magnetic field has no y-component by design. We can thus express the recursion relation in terms of the wavevector magnitude $k$ and the x and z components or the magnetic field. In terms of Hall parameters these are $\chi_x = \frac{e B_x \tau_{ei}}{m_e}$ and $\chi_z = \frac{e B_z \tau_{ei}}{m_e}$ respectively. If we also work in the region of low-frequency waves, we can assume the Fourier transform of the time derivative term in Eq. 20 is ignorable. Then at each order of $l>1$ there is the 5-point recursion relation of the form,
\begin{widetext}
\begin{equation}
    \left[\frac{l(l+1)}{2} - i\chi_x m\right]f^m_l+i \eta\left[\frac{l-m}{2l-1}f^m_{l-1}+\frac{l+m+1}{2l+3}f^m_{l+1}\right]-\frac{\chi_z}{2}\left[ (l-m)(l+m+1)f^{m+1}_l-f^{m-1}_l\right]=0.
\end{equation}
\end{widetext}

Here $\eta= k \lambda_{ei}$ is the non-dimensional Fourier-space Knudsen number utilising the wavevector magnitude $k$. 

\subsection{Matrix Recursion Relation}

This problem can be recast as a matrix recursion relation, with a vector $\textbf{f}_l$ of length $l+1$ with elements $f^0_l, f^1_l, f^2_l, ...$. This recursion relation will have matrix coefficients of size $(l+1)\times(l+1)$. The objective, in analogy with the continued fraction method above, is to close the recursion relation with a matrix that contains contributions from orders $l+1,l+2,l+3 , etc$. This will entail finding a recursion relation closure of the form,
\begin{equation}
\textbf{f}_{l+1}=S_{l+1}\textbf{f}_l,
\end{equation}

such that only the $l=1$ equation is necessary and all the information from the higher modes is contained in the matrix $S_{l+1}$. In this way the influence of the entire infinite set can be included in only the first two orders.

For each order of $l$, Eq. 25 has the matrix form,
\begin{equation}
    M_l\textbf{f}_l+i\eta N_l\textbf{f}_{l-1}+i\eta P_l\textbf{f}_{l+1}=0.
\end{equation}

The matrix $M_l$ is tridiagonal with size $(l+1)\times(l+1)$ and defined as
\[ M_l=
\begin{bmatrix}
    & ... & \\
    \frac{\chi_z}{2}, & \frac{l(l+1)}{2} - i\chi_x m, & -\frac{\chi_z}{2} (l-m)(l+m+1) \\
    & ... & \\
\end{bmatrix}
\]

The matrices $N_l$ and $P_l$ are diagonal, with the diagonal elements (with $m \in [0,l]$)
\begin{align}
    N^m_l &= \frac{l-m}{2l-1}\\
    P^m_l &= \frac{l+m+1}{2l+3}
\end{align}

Using the closure relation it is possible to convert the recurrence relation to the form,
\begin{equation}
   \left( M_l+i\eta  P_l S_{l+1}\right)\textbf{f}_l= - i N_l\textbf{f}_{l-1},
\end{equation}
and by comparison with equation 27, we can find the matrix $S_{l+1}$ defined in terms of a matrix recursion relation,
\begin{equation}
    S_l=-\left( M_l+i \eta P_l S_{l+1}\right)^{-1} i \eta  N_l.
\end{equation}
However it will be useful to introduce a secondary form using the matrix G, defined as,
\begin{equation}
    G_l= M_l+i \eta P_l S_{l+1},
\end{equation}
which follows the matrix recursive definition,
\begin{equation}
     G_l= M_l+\eta^2 P_l G_{l+1}^{-1} N_{l+1}.
\end{equation}
Comparing this equation to equation 15, one can see the analogy with a continued fraction of the form of eq. 16
\begin{align*}
    G_l= M_l+\eta^2 P_l \left(M_{l+1}+\eta^2 P_{l+1} G_{l+2}^{-1} N_{l+2}\right)^{-1} N_{l+1}.
\end{align*}

We can see from this definition that if the plasma is perfectly local, i.e $\eta=0$, the matrix $G_l$ is equal to $M_l$ and there is no need for a recursive definition as there is no coupling to successive orders of $l$.

To close the system at the first order $l=1$ and find Cartesian forms, we make use of the KALOS relations   
\begin{gather}
f_l^x=f_l^0,\\
f_l^y=2\Re(f_l^1),\\
f_l^z=-2\Im(f_l^1),
\end{gather}
to map the elements of $G_1$ to the Cartesian coordinate system by using the mapping 
\begin{equation*}
\textbf{G}^{-1} = \begin{bmatrix}
G_l^{0 0}& G_l^{01}/2 & 0 \\
G_l^{1 0} & G_l^{1 1} & 0 \\
0 & 0 & - G_l^{11}/2
\end{bmatrix}
\end{equation*}

\subsection{Incorporating the $f_0$ equation}

So far, there has been no mention of the $f_0$ equation and the influence it will have on the correction factors. $f_0$ is assumed to be a local Maxwellian in fluid and simple non-local models \cite{Epperlein1986ThePlasma}, however the non-local deviation from a Maxwellian must be included, and as such the $f_0$ equation must form part of the equation set. The linearised $f_0$ equation in Fourier space is
\begin{equation}
    \frac{\partial \delta \hat{f}_0}{\partial t}+i\frac{v}{3}\textbf{k}\cdot\hat{\textbf{f}}_1=C_{e0}[\delta \hat{f}_0].
\end{equation}
For the purpose of qualitatively incorporating e-e collisions, we use a BGK e-e collision operator for $C_{e0}$, the Laplace transform of the above equation yields
\begin{equation}
    s \delta \hat{f}_0-\delta \hat{f}_m(0)+i\frac{v}{3}\textbf{k}\cdot\hat{\textbf{f}}_1=-\nu_{ee}\delta \hat{f}_{nl}.
\end{equation}
Where $\delta \hat{f}_{nl}$ represents the non-local contribution to the deviations from global equilibrium. While the use of a BGK operator is not as exact as the full Fokker-Planck operator for e-e collisions, it is used in other non-local models such as the SNB model \cite{Schurtz2000ACodes}. The tractability of the BGK operator allows an analytic treatment, and we choose to explore the consequences thereof, leaving a more accurate e-e operator for later work.

The initial condition is represented by a perturbed, sinusoidal Maxwellian,
\begin{equation}
    \delta \hat{f}_m(0) = \delta \hat{m}(0)f_m= \left[\frac{\delta \hat{n}(0)}{n_0}+\frac{\delta\hat{T}(0)}{T_0}\frac{1}{2}(\frac{v^2}{v^2_{th}}-3)\right]f_m.
\end{equation}
The linearised $\textbf{f}_1$ equation (eq. 14) similarly is
\begin{equation}
    i\textbf{k}v\delta \hat{f}_0-\hat{\textbf{a}} \frac{v}{v_{th}^2}f_m=-\nu_{ei}\textbf{G}^{-1}\hat{\textbf{f}}_1.
\end{equation}

If the deviation from the global $f_0$ is taken as the sum of local Maxwellian and non-local parts, $\delta f_0 = \delta f_m+\delta f_{nl}$, and we eliminate $\delta f_0$, this leaves
\begin{equation}
    -\nu_{ei}\left[\textbf{K}+\textbf{G}^{-1}\right]\hat{\textbf{f}}_1=\left[i\textbf{k}v\delta \hat{Q}-\hat{\textbf{a}} \frac{v}{v_{th}^2}\right]f_m.
\end{equation}
Here the matrix elements $G_{ij}$ correspond to the Cartesian elements of the matrix $G_1$ using the relations eqs. 34-36. Therefore the terms $G_{ij}$ correspond to the corrections from $l>1$ and the matrix $\textbf{K}$ corresponds to the corrections from the $f_0$ equation. The factor $\delta \tilde{Q}$ is defined as
\begin{equation}
    \delta \hat{Q}=\frac{\delta \hat{m}(0)+\nu_{ee}\delta \hat{m}}{s+\nu_{ee}},
\end{equation}
and the matrix $\textbf{K}$ is defined as
\begin{equation}
    K_{ij}=\frac{k_i k_j v^2}{3\nu_{ei}(s+\nu_{ee})}.
\end{equation}

It is possible to use the quasi-static approximation for $f_0$ in which case we can assume $ \delta \hat{Q}=\delta \hat{m}$. Whilst the value of $s$ is large at high frequency, for a quasi-collisional plasma $s<<\nu_{ee}$, and so $s$ can be dropped without detrimentally effecting the values of the coefficients \cite{Brunner2000LinearTransport}. This also allows us to use the form of the electron mean free path $\lambda_{mfp}= \sqrt{\lambda_{ee}\lambda_{ei}}\approx \sqrt{Z}\lambda_{ei}$.

In this case, the non-local influence of the $f_0$ equation on $\textbf{f}_1$ is held within the matrix $\textbf{K}$. It is possible to invert $\textbf{K}+\textbf{G}^{-1}$ such that both the high-order corrections and the elements of $\textbf{K}$ are held within a 'total' magnetisation matrix $\textbf{G}' = (\textbf{K}+\textbf{G}^{-1})^{-1}$. 

\section{The Corrected Transport Coefficients}
\label{S:4}
We are now in a position to calculate the modified magnetisation matrix $\textbf{G}$. This matrix can then be used in equation 18 in place of the unmodified matrix $\textbf{M}$ in the $\textbf{f}_1$ equation. This matrix is evaluated by calculating the recursive definition of $G_l$ (Eq. 33) as a function of $\eta,v,\chi$ up to an arbitrarily large value of $l$. A numerical calculation is performed up to $l_{max}=200$, for a mesh of values of these parameters across a wide range of scales.

With this $\textbf{G}$ matrix, the transport coefficients in the linear regime can be recalculated using the expressions found in the Appendix Eqs. A23-A31 and the modified moments Eqs. A20-A22, only this time, we use $\textbf{G}$ instead of $\textbf{M}$. The first step will be to compare them to the coefficients calculated by Epperlein and Haines \cite{Epperlein1986PlasmaEquation}. This will show how non-locality and the expansion truncation changes thermal conductivity, resistivity and the thermoelectric effect.

\begin{figure*}
    \centering
    \includegraphics[width=\textwidth]{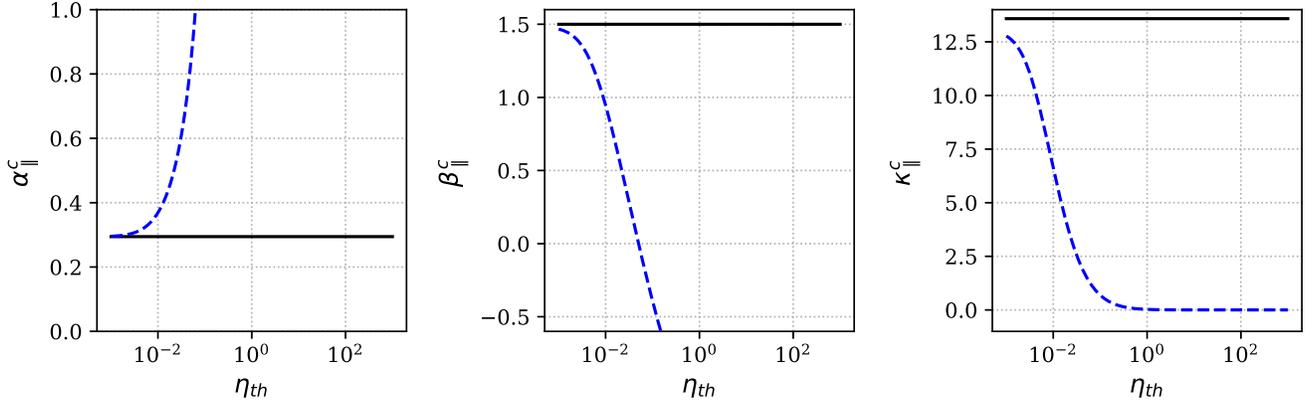}
    \caption[Comparison of the parallel coefficients with the local approximation]{The black lines show the local parallel transport coefficients calculated by Epperlein and Haines in the Lorentz approximation. The non-local results presented in this paper are shown with dashed blue lines. The non-local correction has led to resistivity growing with $\eta$ and thermal conductivity and the thermoelectric term dropping off to zero as $\eta$ increases.}
    \label{fig:paracompare}
\end{figure*}

\begin{figure}
    \centering
    \includegraphics[width=\columnwidth]{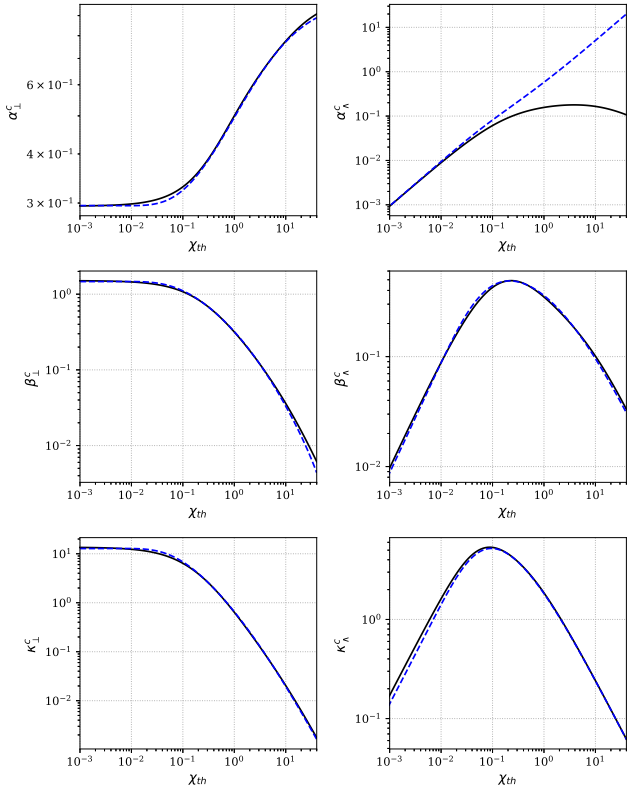}
    \caption[Comparison of the magnetised coefficients with the local approximation]{Comparison of the perpendicular and cross-perpendicular transport coefficients (blue dashed lines) presented in this paper, with the Epperlein and Haines results (solid black lines) in the Lorentz limit.}
    \label{fig:magcompare}
\end{figure}

Figures \ref{fig:paracompare} and \ref{fig:magcompare} show the comparison between the Epperlein and Haines coefficients \cite{Epperlein1986PlasmaEquation}  and the new corrected forms. Figure \ref{fig:paracompare} plots the normalised parallel coefficients against non-locality and reassuringly as the non-locality parameter $\eta\rightarrow 0$ the new coefficients converge on the Epperlein and Haines values in the Lorentz limit ($Z\rightarrow\infty$). Figure \ref{fig:magcompare} focuses on the perpendicular and cross-perpendicular coefficients in the local approximation. Again we see a reassuring equivalence with the Epperlein and Haines dependence on the thermal Hall parameter $\chi_{th}=\chi(v_{th})$. This is apart from the highly magnetised cross-perpendicular resistivity $\alpha_\wedge$ where it peaks at a much higher value. This divergence from the result of Epperlein and Haines was also seen in previous work \cite{Epperlein1986ThePlasma}. While the values of $\alpha_\wedge$ converge asymptotically, the intermediate deviation corresponds not to non-locality but to the difference between the fitted form of \cite{Epperlein1986PlasmaEquation} and the calculation in \cite{Epperlein1986ThePlasma}.    

\begin{figure}
    \centering
    \includegraphics[width=\columnwidth]{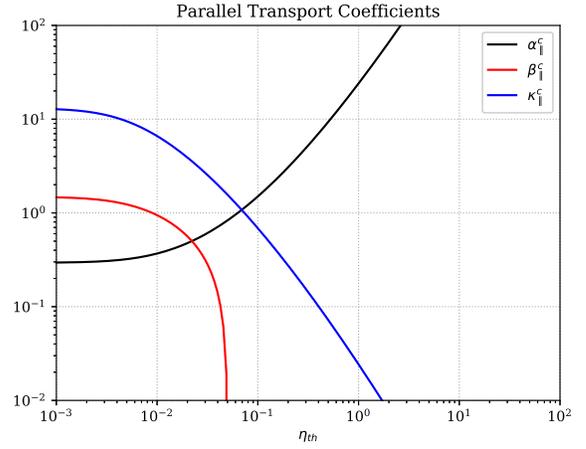}
    \caption[The parallel transport coefficients]{The parallel transport coefficients as a function of non-locality $\eta_{th}$. $\beta_\parallel$ drops quickly with $\eta_{th}$ before changing sign, while $\alpha_\parallel$ and $\kappa_\parallel$ asymptotically have a linear depenence on $\eta_{th}$ }
    \label{fig:paraline}
\end{figure}

Figure \ref{fig:paraline} plots the parallel coefficients against non-locality parameter. Given there is no dependence on magnetisation for the parallel terms, the dependence on $\eta_{th}=\eta(v_{th})$ is monotonic. The parallel thermal conductivity and thermoelectric coefficient decrease, with $\beta_\parallel$ dropping sharply at $\eta_{th}\approx 0.05$. The resistivity mirrors the thermal conductivity with a steady increase with increasing non-locality.

\begin{figure}
    \centering
    \includegraphics[width=\columnwidth]{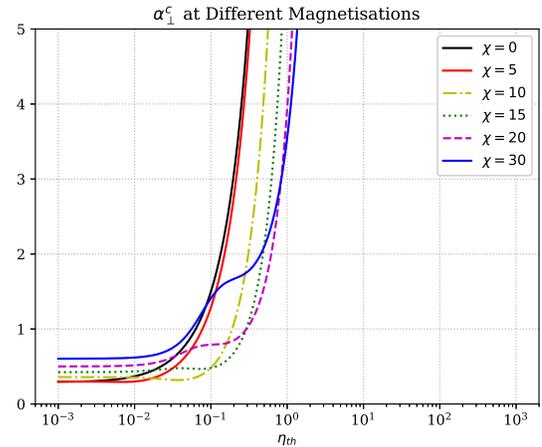}
    \caption[Lineouts of $\alpha_\perp$]{These lineouts of $\alpha_\perp$ for different magnetisation values show an inflection point at $\eta_{th} \approx 0.1$ when $\chi_{th}>15$}
    \label{fig:aperpline}
\end{figure}

\begin{figure}
    \centering
    \includegraphics[width=\columnwidth]{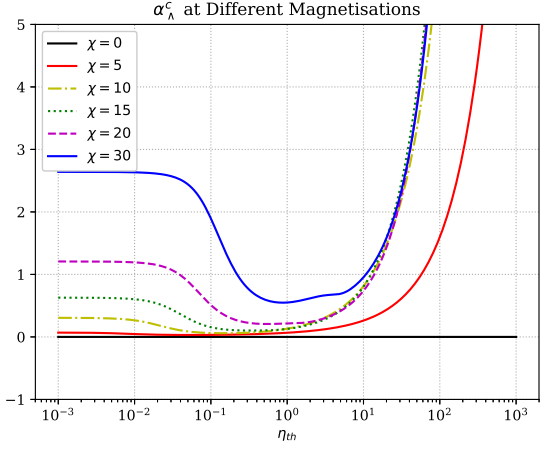}
    \caption[Lineouts of $\alpha_\wedge$]{When $\alpha_\wedge$ is magnetised, a drop appears at $\eta_{th}\approx 0.05$, with the depth of the valley larger at higher values of $\chi_{th}$}
    \label{fig:awedgeline}
\end{figure}

\begin{figure}
    \centering
    \includegraphics[width=\columnwidth]{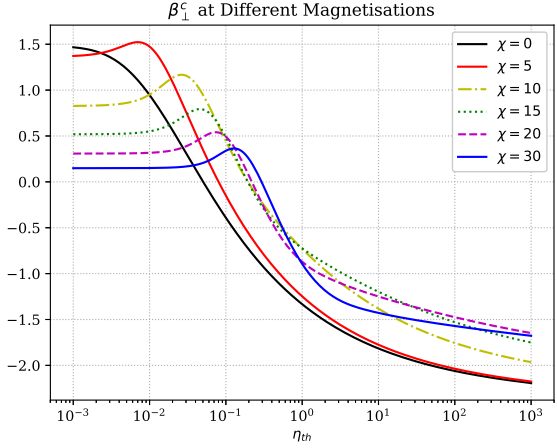}
    \caption[Lineouts of $\beta_\perp$]{Peaks in $\beta_\perp$ appear when $\chi_{th}>0$, before eventually changing sign at a value of $\eta_{th}$ that depends on the value of $\chi_{th}$}
    \label{fig:bperpline}
\end{figure}

\begin{figure}
    \centering
    \includegraphics[width=\columnwidth]{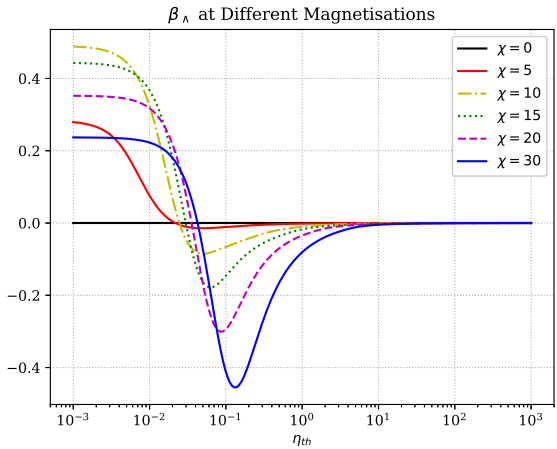}
    \caption[Lineouts of $\beta_\wedge$]{Lineouts of $\beta_\wedge$ tend to zero at very high $\eta_{th}$ at all values of $\chi_{th}$. The swing from positive to negative is more extreme at higher $\chi_{th}$, with the maximum, minimum and zero all dependent on $\chi_{th}$}
    \label{fig:bwedgeline}
\end{figure}

\begin{figure}
    \centering
    \includegraphics[width=\columnwidth]{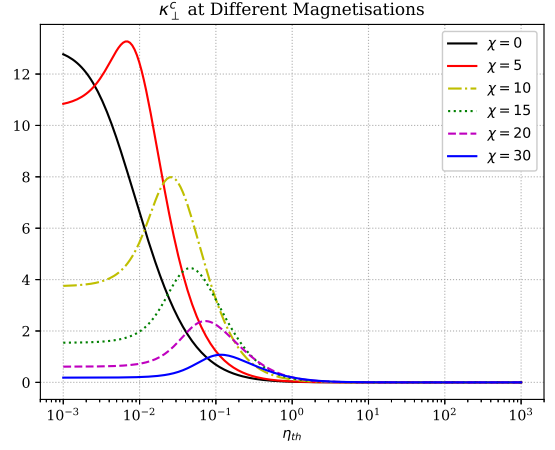}
    \caption[Lineouts of $\kappa_\perp$]{Lineouts of $\kappa_\perp$ show a peak with a value up to twice the local value, with a position that shifts to the right as magnetisation increases.}
    \label{fig:kperpline}
\end{figure}

\begin{figure}
    \centering
    \includegraphics[width=\columnwidth]{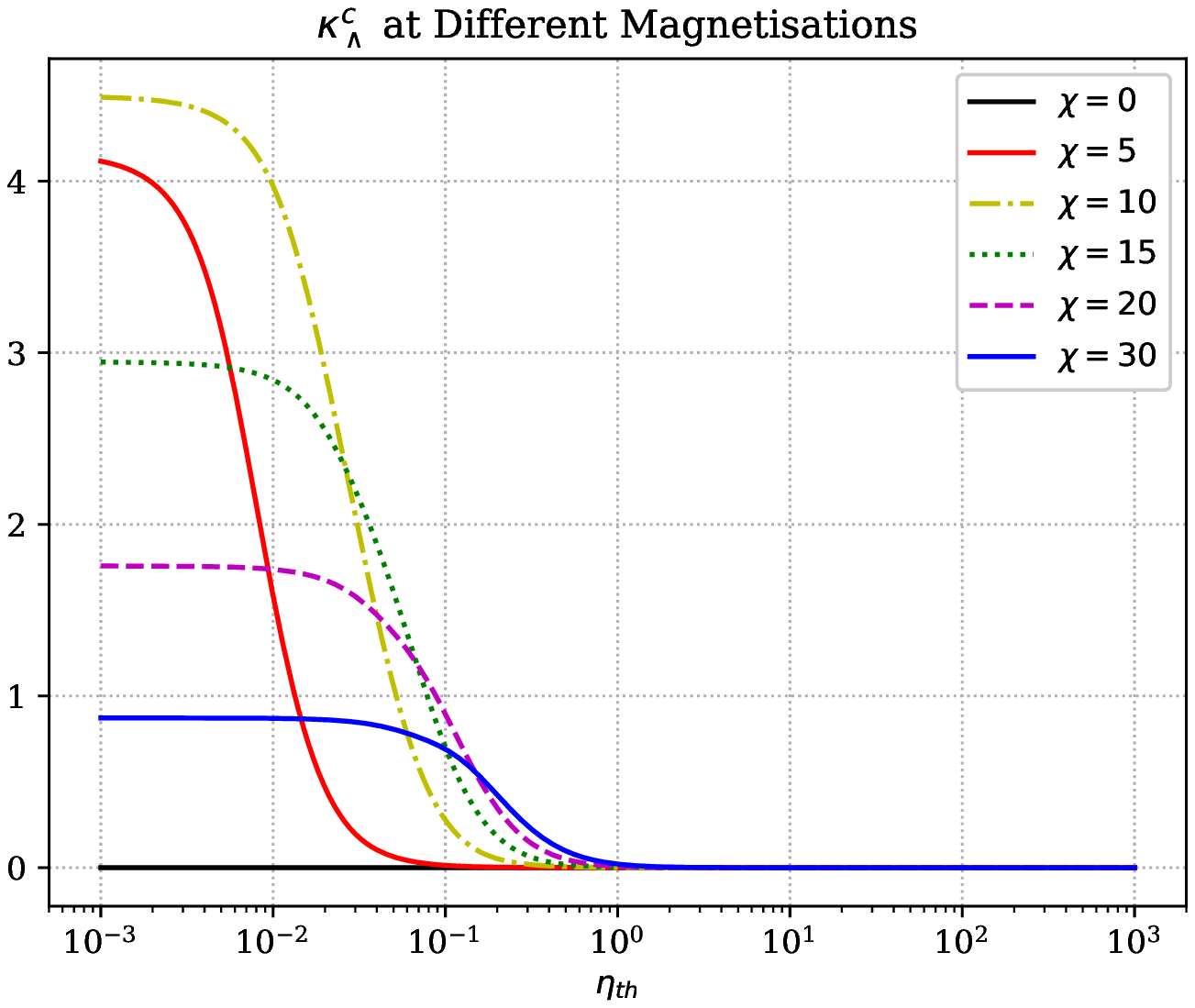}
    \caption[Lineouts of $\kappa_\wedge$]{Though $\kappa_\wedge$ vanishes at high $\eta_{th}$ for all magnetisation values, the drop off shifts to the right as magnetisation increases.}
    \label{fig:kwedgeline}
\end{figure}

The perpendicular and cross-perpendicular resistivity, shown in figures \ref{fig:aperpline} and \ref{fig:awedgeline}, reveal the signs of the complex interplay between magnetisation and non-locality. The most significant feature of the magnetised non-local coefficients in this linear regime is shared by both the thermal conductivity and thermoelectric coefficients. All four show a peak, with a value greater than the local ($\eta_{th}=0$) value, which appears and grows as $\chi_{th}$ increases, illustrated in lineouts figures \ref{fig:bperpline}, \ref{fig:bwedgeline},\ref{fig:kperpline} and \ref{fig:kwedgeline}. This in turn shifts the curves to higher values of $\eta_{th}$. This non-monotonic dependence appears as magnetisation grows.

Given the diffusive approximation corresponds to ignoring all terms with $l>1$, an important question is how the infinite continued fraction differs from this truncated version. How do the correction factors change as higher orders are included and does the result converge? Looking at the convergence of these functions, it is clear there is significant deviation at intermediate values of non-locality. This is entirely the result of higher-order modes that are ignored in almost all closures and transport models. 

\section{Implications in the context of laser-plasma experiments}
\label{S:5}

The explanation for this non-monotonic dependence on non-locality of the transport coefficients lies in the coupling between different directions in a magnetised plasma. The magnetic field term in the VFP couples together the $x, y$ components. When this is repeated in the recursion relation there is a coupling from $x$ to $y$ and back again $\textit{ad infinitum}$. The localising nature of magnetisation perpendicular to the direction of the magnetic field leads to a minimum in $H_\perp$ and the corresponding non-monotonic behaviour with non-locality in the transport coefficients.   

In particular, the reversal of the $\beta$ coefficients corresponds to the value of $\eta_{th}$ where $\langle\textbf{M}\tilde{v}^9\rangle=5\langle\textbf{M}\tilde{v}^7\rangle$ in Eq. A5. As the moments $\langle\tilde{v}^n\rangle$ are functions of the non-locality parameter $\eta_{th}$, there is a point where the thermoelectric term in the Onsager forms cancels out. Whilst the reversal of the $\beta$ sign has been reported before in unmagnetised plasmas \cite{Brantov1996NonlocalHydrodynamics}, it has not been seen for the full magnetised transport coefficients. 

This also implies the Nernst effect switches direction at high values of $\eta_{th}$, since the thermal conductivity $\kappa$ does not reverse in a similar manner, there is advection against heat flow. One can consider the Nernst effect acting on a single temperature mode $\delta T$. As the wavenumber of the mode increases, we can see from figure 13 that at some point - dependent on the value of magnetisation - the Nernst effect will act in anti-phase to the thermal diffusion. In doing so the Nernst effect no longer approximately follows the heat flow as used by Haines \cite{Haines_1986}.

In the context of laser-plasma experiments, experiments have been performed in both ICF-like \cite{Froula2007} and MAGLIF schemes \cite{Slutz2016ScalingAccelerators} where the magnetisation parameter $\chi$ is of the order $>10$. Comparing the temperature and density scales and the expected non-locality, magnetised plasma physics experiments frequently enter regimes where the transport coefficients differ by up to a factor of 2 because of the effects described above. In addition, if the Nernst (driven by $\beta_\wedge$) and Cross-Nernst (driven by $\beta_\perp$) are as significant as recent simulations suggest \cite{Walsh2017Self-GeneratedFacility, Hill2018EnhancementConditions}, their reversal of direction will further change the dynamics away from what current simulations suggest.  

\section{Conclusions}
\label{S:6}

In summary, in the linear regime the influence of magnetised non-locality can be aggregated into three correction functions $H_\parallel, H_\perp, H_\wedge$. This is performed using a matrix recursion method to arbitrarily high orders of the spherical harmonic expansion. These corrections represent the non-locality parallel, perpendicular and cross-perpendicular to the applied magnetic field respectively. Using these corrections to derive a set of transport coefficients, we can show these coefficients differ from the classical approximation. The unexpected non-monotonic dependence on non-locality as these coefficients become increasingly magnetised has not before been reported. In turn, the peaks with up to double the diffusion approximation value at intermediate values of non-locality, would lead to as-yet unseen transport phenomena in the linear regime.

Though this work looked only at the case of a magnetic field perpendicular to the wavevector of a perturbation, these results could be extended to an arbitrarily-directed magnetic field. In this case, the recurrence relation would include more terms and the $H$ function corrections would include contributions from all elements of the matrix $G$. We also only considered plasmas in the high ion charge regime ($Z>>1$). Further future work will include the impact of a more accurate electron-electron collision physics.    

By incorporating the higher-order modes into transport calculations, we present a possible source of error in simulations of plasmas under magnetic fields. This improves our understanding of magnetised transport and can be used directly in the analysis of damping of waves in plasmas. Overall, this result provides motivation for the closer study of the terms ignored in most fluids closures. We conclude it is necessary to use higher-order terms of the spherical harmonic expansion when constructing a closure in magnetised plasmas. 

Inertial fusion experiments today work in regimes where the transport in the plasma is non-local. With the use of very strong magnetic fields, a better understanding of the interplay between non-locality and magnetisation is required. As we have demonstrated in this work, the deviation from classical transport can change the expected transport of charge, heat and magnetic fields in a plasma.

\begin{acknowledgments}
The authors thank S. Mijin and Dr D. Hill for helpful comments on this work. This work was funded through ESPRC doctoral training grant (No. EP/M507878/1) and via an AWE plc CASE partnership.
\end{acknowledgments}

\section{Data Availability}
Data sharing is not applicable to this article as no new data were created or analyzed in this study.

\appendix

\section{Calculating the Local Transport Coefficients}
\label{S:Ap}

The electric field that corresponds to the electron pressure can be separated out of the $ \delta \hat{f}_0$ term by considering the form of Ohm's law. Denoting this field by $e n_e\textbf{E}' = -\nabla p_e$ and using the ideal gas equation of state, in Fourier space it becomes
\begin{equation}
    \hat{\textbf{E}}'=i\textbf{k}\frac{T_0}{e}\left(\frac{\delta\hat{T}}{T_0}+\frac{\delta \hat{n}}{n_0}\right),
\end{equation}
from which we define
\begin{equation}
    \hat{\textbf{a}}'=-\frac{e\hat{\textbf{E}}'}{m_e}=-i\textbf{k}\frac{T_0}{m_e}\left(\frac{\delta\hat{T}}{T_0}+\frac{\delta \hat{n}}{n_0}\right).
\end{equation}
The total electric field will now be called $\textbf{a}^*=\textbf{a}'+\textbf{a}$. 

To simplify the analysis, the velocity is re-parameterised such that $\tilde{v}=v/v_{th}$ and the equation becomes,
\begin{equation}
    \hat{\textbf{f}}_1=-\frac{\tilde{v}^4}{\nu_{T}v_{th}} \textbf{M} \left[i \textbf{k} \frac{\delta \hat{T}}{m_e}\frac{1}{2}(\tilde{v}^2-5)-\textbf{a}^*\right]f_m,
\end{equation}
where $\nu_T=\nu_{ei}(v_{th})$. In order to simplify the integration procedure whilst taking moments of this equation, we introduce the integral
\begin{equation}
    \frac{n_0}{v_{th}^3}\langle\textbf{M}\tilde{v}^n\rangle= 4\pi\int^\infty_0 \textbf{M}\tilde{v}^n f_m d\tilde{v}.
\end{equation}
The magnetisation matrix moment integral $\langle\textbf{M}\tilde{v}^n\rangle$ can be written out in full using the definition of the magnetisation matrix (eq. 9),
\begin{multline*}
\langle\textbf{M}\tilde{v}^n\rangle = 4\pi\int^\infty_0 \frac{\chi_i \chi_j}{\chi^2}\tilde{v}^n f_m d\tilde{v} \\+ 4\pi\int^\infty_0 \left(\delta_{ij}-\frac{\chi_i \chi_j}{\chi^2}\right)\frac{1}{1+\chi^2}\tilde{v}^n f_m d\tilde{v}\\- 4\pi\int^\infty_0\frac{\epsilon_{ikj}\chi_k}{1+\chi^2}\tilde{v}^n f_m d\tilde{v},
\end{multline*}
with $\tilde{f}_m=(2\pi)^{3/2}\exp{-\tilde{v}^2/2}$  and $\chi_{th}=\chi(v_{th})$. In the analysis of the body of this paper this expression is evaluated numerically.

If this expression is now used alongside the moment definitions of heat flow and current, eqs. 11 and 12, these fluxes can be written,
\begin{widetext}
\begin{gather}
    \hat{\textbf{j}}\left(- \frac{3}{ e v_{th}^4}\right)=\frac{n_0}{\nu_{T} v_{th}^4}\left[\langle\textbf{M}\tilde{v}^7\rangle\textbf{a}^*-i \textbf{k} \frac{\delta \hat{T}}{m_e}\frac{1}{2}\left(\langle\textbf{M}\tilde{v}^9\rangle-5\langle\textbf{M}\tilde{v}^7\rangle\right)\right]\\
    \hat{\textbf{q}}\left(\frac{6}{m_e v_{th}^6}\right)=\frac{n_0}{\nu_{T} v_{th}^4}\left[\langle\textbf{M}\tilde{v}^9\rangle\textbf{a}^*-i \textbf{k} \frac{\delta \hat{T}}{m_e}\frac{1}{2}\left(\langle\textbf{M}\tilde{v}^{11}\rangle-5\langle\textbf{M}\tilde{v}^9\rangle\right)\right].
\end{gather}
\end{widetext}

This transport pair can now be put into the Onsager \cite{Helander2005CollisionalPlasmas} form with matrix-valued coefficients $\bm{\rho}, \bm{\sigma},\bm{\zeta},\bm{\mu}$,
\begin{gather}
    \hat{\textbf{j}}=\bm{\sigma}\textbf{E}^*+i\bm{\rho}\textbf{k}\delta \hat{T}\\
    \hat{\textbf{q}}=i\bm{\zeta}\textbf{k}\delta \hat{T}+\bm{\mu}\textbf{E}^*.
\end{gather}

In this form each thermodynamic flux ($\textbf{j},\textbf{q}$) is driven by thermodynamic forces ($\textbf{E}, \delta T$). It will however be of more use to use the classical transport forms used in the literature. Thus we invert the current equation eq. A7 and re-express the heat flow equation in terms of the intrinsic heat flow $\textbf{q}'$, following Bychenkov  \cite{Bychenkov1995NonlocalPlasma}.
\begin{gather}
    e n_0\textbf{E}^*=\bm{\alpha}\textbf{j} -n_0 i\bm{\beta}\textbf{k}\delta \hat{T}  \\
    \hat{\textbf{q}}'=-i\bm{\kappa} \textbf{k}\delta \hat{T} -\bm{\beta} \textbf{j} \frac{T_0}{e}
\end{gather}

In this form we consider the thermal conductivity $\bm{\kappa}$, the resistivity $\bm{\alpha}$ and the thermoelectric coefficient $\bm{\beta}$. By comparison with the Onsager forms, we can put the classical coefficients in terms of the Onsager coefficients,
\begin{gather}
    \bm{\alpha}=en_0\bm{\sigma}^{-1}\\
    \bm{\kappa}=-(\bm{\zeta}+\bm{\mu}\bm{\sigma}^{-1}\bm{\rho})\\
    \bm{\beta}=-e\bm{\sigma}^{-1}\bm{\rho}.
\end{gather}
By collecting the terms of the above equations, we can see the coefficients have the form,
\begin{gather}
    \bm{\alpha}=\frac{3 m_e \nu_{T}}{e}\langle\textbf{M}\tilde{v}^7\rangle^{-1},\\
    \bm{\kappa}=\frac{n_0 v^2_{th}}{\nu_{T} 12}\left[\langle\textbf{M}\tilde{v}^{11}\rangle-\langle\textbf{M}\tilde{v}^9\rangle\langle\textbf{M}\tilde{v}^7\rangle^{-1}\langle\textbf{M}\tilde{v}^9\rangle\right],\\
    \bm{\beta}=\frac{1}{2}\left[ \langle\textbf{M}\tilde{v}^7\rangle^{-1}\langle\textbf{M}\tilde{v}^9\rangle-5\textbf{I}\right].
\end{gather}These transport coefficients are conventionally expressed in terms of the parallel ($\parallel$), perpendicular ($\perp$) and cross-perpendicular ($\wedge$) to the magnetic field. A general transport coefficient $\bm{ \eta }$ with a driving gradient  $\textbf{s}$ can be expressed as

\begin{equation}
\bm{ \eta }\cdot\textbf{s} =\eta_\parallel (\hat{\textbf{b}}\cdot \textbf{s}) + \eta_\perp \hat{\textbf{b}} \times ( \textbf{s} \times \hat{\textbf{b}}) + \eta_\wedge \hat{\textbf{b}} \times \textbf{s},
\end{equation}
with $\hat{\textbf{b}} = \frac{\textbf{B}}{|\textbf{B}|}$. From this expression one can find the parallel, perpendicular and cross-perpendicular expressions from the cartesian form of the transport coefficient. In particular, if one chooses a geometry where $\textbf{B} = (B_x, 0, B_z)$ and $\textbf{s} = (s_x, 0, 0)$ as is the case in \ref{S:3} the form of the parallel, perpendicular and cross-perpendicular coefficients are
\begin{gather}
\eta_\parallel = \frac{b_x \eta_{xx} + b_z \eta_{zx}}{b_x^3 + b_x b_z^2},\\
\eta_\perp = \frac{b_z \eta_{xx} - b_x \eta_{zx}}{b_x^2 b_z + b_z^3},\\
 \eta_\wedge = \frac{\eta_{yx}}{b_z}.\\
\end{gather}

\bibliographystyle{unsrt}
\bibliography{references.bib}

\end{document}